\documentclass[prd,twocolumn,nofootinbib,10pt,aps]{revtex4-1}
\date{\today}

\usepackage{epsfig}
\usepackage{amsmath}
\usepackage{amsfonts}
\usepackage{graphicx}
\usepackage{dsfont}
\usepackage[german, english]{babel}
\usepackage{amssymb}
\usepackage{ifthen}
\usepackage{ulem}
\usepackage{color}

\RequirePackage[colorlinks=true,allcolors=blue,
unicode=true,hypertexnames=false]{hyperref}
\hypersetup{pdfstartview={XYZ null null 1.25}}

\begin{document}
	
	\title{Synthetic nuclear Skyrme matter in imbalanced Fermi superfluids with a multicomponent order parameter}
	\author{Albert Samoilenka}
	\affiliation{Department of Physics, KTH Royal Institute of Technology, 106 91 Stockholm, Sweden}
	
	\author{Filipp N. Rybakov}
	\affiliation{Department of Physics, KTH Royal Institute of Technology, 106 91 Stockholm, Sweden}
	
	\author{Egor Babaev}
	\affiliation{Department of Physics, KTH Royal Institute of Technology, 106 91 Stockholm, Sweden}

	\begin{abstract}
		Cooper-pair formation in a system of imbalanced fermions leads to the well-studied Fulde-Ferrell or Larkin-Ovchinnikov superfluid state. In the former case the system forms spontaneous phase gradients while in the latter case it forms a stripelike or a crystal-like density gradient. We show that in multicomponent imbalanced mixtures, the superfluid states can be very different from the Fulde-Ferrell-Larkin-Ovchinnikov states. The system generates gradients in both densities and phases by forming three-dimensional vortex-antivortex lattices or lattices of linked vortex loops. The solutions share some properties with the ostensibly unrelated Skyrme model of densely packed baryons and can be viewed as synthetic realization of nuclear Skyrme matter.
	\end{abstract}
	
	\maketitle
	
	\section{Introduction}
	
	One of the most important examples of translation symmetry breaking 
	in condensed matter states is formation of a crystal of
	topological defects.  
	In the simplest superconductors and superfluids, the topological
	defects originate from the fact that they are described
	by a complex field $\Psi=|\Psi|e^{i\theta}$, where the $2\pi$-periodic field $\theta$ is called superfluid phase.
	As a consequence superfluid and superconducting vortices
	are characterized by one of the simplest
	topological
	indexes: the integer phase winding $N$.  It quantifies how many times the phase changes
	from $0$ to $2\pi$ when one circumvents a vortex core.  It is given by the line integral $N = \frac{1}{2 \pi} \oint \nabla \theta \cdot \vec{dl} $.
	One of the state-defining properties
	of superfluids is that it forms vortex lattices under rotation \cite{Onsager:49,Feynman:55}.
	Superconductors form vortex lattices in an external magnetic field \cite{Abrikosov:57}.
	More recently, lattices of more complicated topological defects, skyrmions, became of paramount importance in magnetism~\cite{Nagaosa_review}.
	The magnetic skyrmion is a two-dimensional texture of a dimensionless magnetization vector $\mathbf{m}(\mathbf{r})$ with a non-zero integer topological invariant given by the integral $q\!=\!\frac{1}{4\pi}\int\varepsilon_{a b c} m_a \partial_x m_b \partial_y m_c dx dy$, where $\varepsilon_{a b c}$ is the Levi-Civita symbol.
	In films, both vortex and skyrmion lattices can be viewed as crystals of two-dimensional particle-like objects, while in three dimensions they can be viewed as lattices of strings-like objects. 
	Consequences of spontaneous breakdown of translation symmetry dictate unique macroscopic responses of these condensed matter systems such as transport, magnetic and thermodynamic properties.

	While  two-dimensional lattices of topological  defects are ubiquitous, the situation is very different in three dimensions. 
	The most celebrated example where three-dimensional crystals of topological solitons were sought is the Skyrme model of nuclear matter~\cite{Skyrme,klebanov1985nuclear,goldhaber1987maximal, kugler1988new, castillejo1989dense}.
	This model  was proposed to describe atomic nuclei as continuous particle-like solutions of the nonlinear field theory in three dimensions, in contrast to two-dimensional magnetic skyrmions.
	
	The nuclear Skyrme model was originally defined in terms of a $SU(2)$ valued chiral field.
	Hence solutions can be characterized by a pair of complex fields $\Phi_1$, $\Phi_2$ constrained by $|\Phi_1|^2 + |\Phi_2|^2 \!=\! 1$.
	Nuclear skyrmion solutions of this model are isolated particle-like states characterized by an integer topological charge (or index) $Q$, which is interpreted as a baryon number and is defined as the integral over the topological density 
	\begin{equation}\label{charge}
	\rho_Q = \frac{1}{12\pi^2} \varepsilon_{abcd} \varphi_a \varepsilon_{ijk} (\partial_i \varphi_b \partial_j \varphi_c \partial_k \varphi_d),
	\end{equation}
	where $\varphi_1\!=\!\mathrm{Re}(\Phi_1)$, $\varphi_2\!=\!\mathrm{Im}(\Phi_1)$, $\varphi_3\!=\!\mathrm{Re}(\Phi_2)$, $\varphi_4\!=\!\mathrm{Im}(\Phi_2)$.
	If one associates vorticity to  a complex field component, then the internal structure of the skyrmion can be interprerted as a bundle of linked closed vortex loops and the linking number corresponds to the value of the topological charge.
	Skyrmions with different topological charges belong to different homotopy classes and therefore cannot mutually transform into each other as a result of any perturbations.
	Particles with charges of the opposite sign attract each other and annihilate, while those with the same sign form highly-charged skyrmions with a morphology resembling Platonic solids~\cite{Atiyah_Sutcliffe}.
	As a consequence it is expected that for an unlimited number of particles with charges of the same sign the ground state is a skyrmion crystal, that is, the nuclear Skyrme matter~\cite{klebanov1985nuclear, goldhaber1987maximal, kugler1988new, castillejo1989dense}.
	Such solutions, although not realized in a terrestrial laboratory, give an important example of a three-dimensional crystal of topological defects, that was hypothesized to share properties with nuclear matter in a neutron star.

	The problem of spontaneous breakdown  of translation symmetry in superfluid  and superconducting states 
	has been of great interest even without formation of topological defects.
	It takes place in Fulde-Ferrell-Larkin-Ovchnnikov (FFLO) states \cite{FF,LO,Bowers_Rajagopal,radzihovsky2010imbalanced,Matsuda2007} which can be viewed as one of the first discussed example of a supersolid state.
	They form in fermionic systems where there is density imbalance of two species of fermions. 
	As a consequence Cooper pairing involves two  fermions with momenta of different magnitude, ensuing a periodic modulation of the phase or modulus of the order parameter.
	In most general considered cases the modulation of the order parameter is two or three dimensional 
	\cite{FF,LO,Bowers_Rajagopal,radzihovsky2010imbalanced,Matsuda2007}.
	In the context of color superconductivity in dense quark matter, three-dimensional modulation is called FFLO crystal~\cite{Bowers_Rajagopal}.
	Such crystalline order is associated with density modulation and is not characterized by topological indices, i.e. 
	is not a lattice of topological defects. While in ultracold atoms the realization of FFLO state in three dimensional 
	parabolic trap was
	challenging, some of the challenges were removed by the invention of the box trap potential ~\cite{box1,box2,box3}.
	Furthermore it was shown that  FFLO-type solutions are much more robust
	near the box potential wall \cite{barkman2019surface,samoilenka2019superconductivity}. 
	Furthermore, since often the phases which are fragile in single-component systems, occupy much larger domain
	in multicomponent case, this raises the question of possible states in multicomponent imbalanced mixtures.
	
	\section{The model}
	
	In this paper we consider mixtures of two species of fermionic ultracold atoms.
	We considered the case where both spin populations are imbalanced. 
	We show that, by contrast, in a two-component system entirely different states arise.
	The system forms textures in the form of  three-dimensional lattice of nuclear skyrmions. This represents  synthetic realization  of the putative Skyrme nuclear matter.
	
	The Ginzburg-Landau functional  for free energy density  of a single-component  FFLO superfluid in BCS limit of weak coupling was derived microscopically in \cite{Radzihovsky}. 
	Its two-component generalization can be written in dimensionless units as
	
	\begin{equation}\label{ham}
	\begin{split}
	\mathcal{F} = \sum_{a = 1}^{2} \bigg( \zeta_a |\Delta\Psi_a|^2 + K_a|\nabla \Psi_a|^2  + \alpha_a |\Psi_a|^2 + \beta_a |\Psi_a|^4  + \\ +
	\xi_a \left(\mathrm{Im}[\Psi_a^*  \nabla \Psi_a]\right)^2 \bigg)  + \gamma|\Psi_1|^2 |\Psi_2|^2,
	\end{split}
	\end{equation}
	
	where $\Psi_1\!=\!f_{1}+i f_{2}$  and  $\Psi_2\!=\!f_{3}+i f_{4}$ is a pair of complex order parameters, $\Psi_a\!=\!|\Psi_a|e^{i\theta_a}$.
	All the microscopic parameters of the model are absorbed in dimensionless coefficients $\zeta_{1,2}$, $K_{1,2}$, $\alpha_{1,2}$, $\beta_{1,2}$, $\xi_{1,2}$.
	Microscopic derivation of parameters in single component case \cite{Radzihovsky} shows that all of them are of order unity (see Appendix).
	Hence the majority of the simulations we performed for coefficients $\zeta_{1,2}\!=\!1$, $K_{1,2}\!=\!-1$, $\alpha_{1,2}\!=\!-1$, $\beta_{1,2}\!=\!1$, $\xi_{1,2} = 1$. 
	In cases where a different value was used, we give the corresponding number.
	The dimensionless unit of length corresponds to $\frac{\hbar v_F}{(\delta\mu)_{c2}}$, where $v_F$ is Fermi velocity and $(\delta\mu)_{c2}$ is critical value for the chemical potential difference $\delta\mu\!=\!|\mu_\uparrow - \mu_\downarrow|$.
	In two-component case one necessarily has to include  additional parameter $\gamma$ that controls the strength of bi-quadratic coupling. 
	In what follows we set it $\gamma \!=\! 0.5$ so that the system is substantially far away from a phase separation regime.
	
	The key feature of systems with FFLO-type instability is that 
	for sufficiently large fermionic imbalance, the coefficient $K$ for the gradient terms becomes negative. 
	Therefore the system forms gradients of the field in the ground state and those should be balanced by retaining positive terms 
	arising at the next order with more spatial derivatives. Note, that a regime is also possible where the coefficient in front of fourth order potential term becomes negative and one should retain sixth-order potential term. While in the recent microscopic Ginzburg-Landau derivation \cite{Radzihovsky} the fourth order potential term does not change sign simultaneously with the second order gradient term, we obtained stable solutions, of the kind discussed below, also in the model with the negative forth- and positive sixth-order potential terms.
	
	\section{The results}
	
\begin{figure*}
		\centering
		\includegraphics[width=0.99\linewidth]{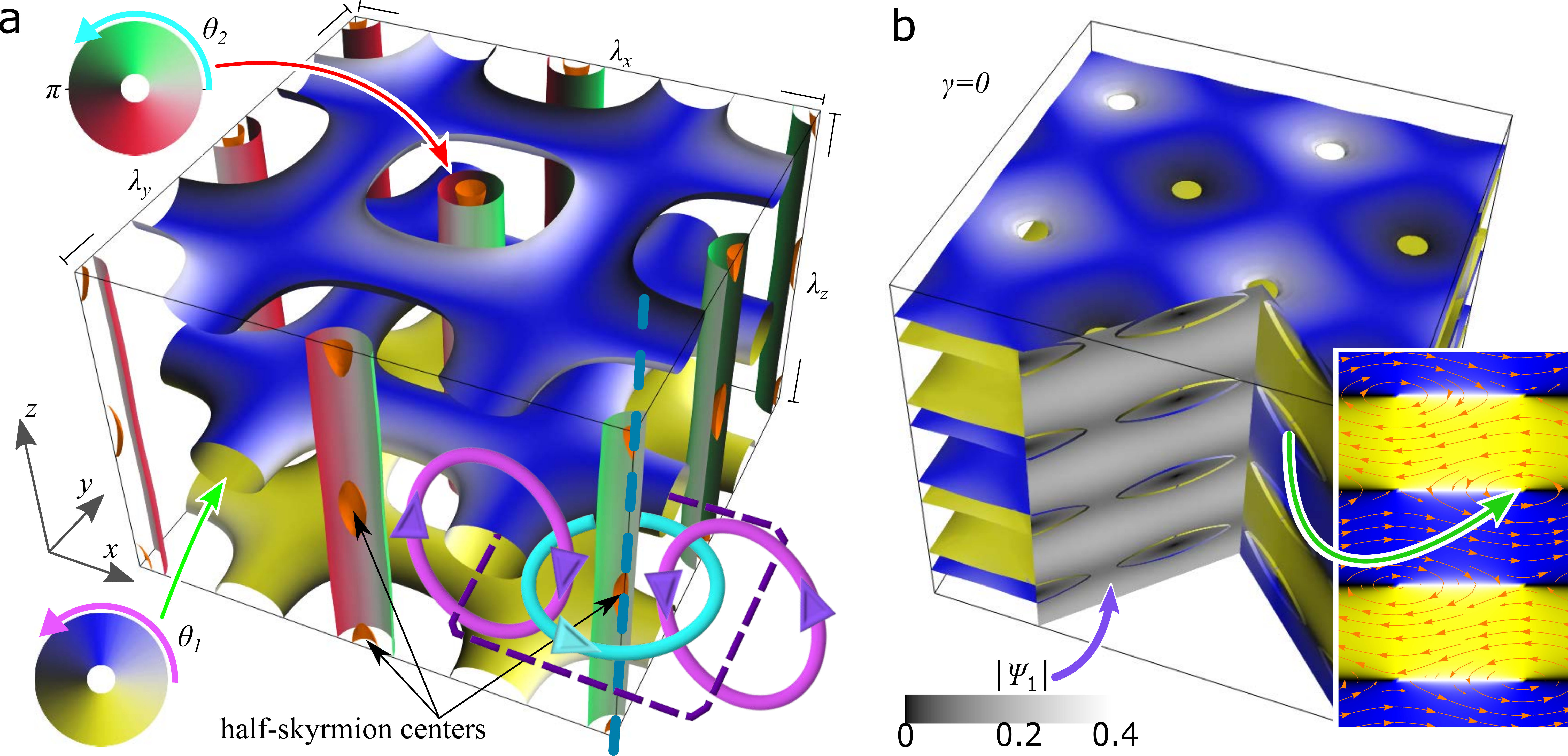}
		\caption{
			The synthetic nuclear skyrmion crystal state in a mixture of imbalanced fermionic superfluids with two-component order parameter. It can be viewed as
			a stable state of linked vortex loops in two complex fields. 
			\textbf{a} The panel shows isosurfaces of order parameters modulus for $|\Psi_{1,2}|\!=\!0.5\,\textrm{max}(|\Psi_{1,2}(\textbf{r})|)$
			in a unit cell of the crystal.
			The color of the isosurfaces shows the values of the corresponding phases $\theta_{1,2}$ in accordance to colormaps in the inserted circles.
			Orange areas show the isosurfaces of topological density for $\rho_Q\!=\!0.95\,\textrm{max}(\rho_Q(\textbf{r}))$.
			Cyan vertical dashed curve denotes the position of one particular vortex core for the second condensate, $\Psi_{2}\!=\!0$.
			Violet dashed curve denotes the position of one particular vortex core for the first condensate, $\Psi_{1}\!=\!0$, which forms
			stable vortex loops.
			Thick circles show several separate closed loops, along which there is a $2\pi$ phase windings in the condensates: cyan corresponds to  $|\Psi_{2}|\!=\!\textrm{const}$, magenta corresponds to $|\Psi_{1}|\!=\!\textrm{const}$.
			The equilibrium periods which were found are: $\lambda_{x,y}\!=\!17.5$, $\lambda_z\!=\!12.5$. Average energy density $\langle\mathcal{F}(\mathrm{\textbf{r}})\rangle\!=\!-0.43$.
			Simulation was performed for $K_{1,2}\!=\!-0.9$.
			\textbf{b} Knackebrod state of uncoupled condensates.
			Isosurfaces and their coloring are the same as in (\textbf{a}).
			The left vertical side of the triangular cut-out shows the density distribution of the condensate in a black and white color code.
			The inset shows the map of streamlines.
			The numerical calculation was performed for $\alpha_{1,2}\!=\!0$.
		}
		\label{SwedBread}
	\end{figure*}

	We investigate numerically stable states of the model~(\ref{ham}) (for details see Appendix). We find that these are three-dimensional crystals with the field configurations $\Psi_{1,2}(\mathbf{r})$ analogous to that for pair $\Phi_{1,2}(\mathbf{r})$ in nuclear skyrmion crystals.
	Thus, the ultra-cold atomic multicomponent mixtures can exhibit the physical realization of a crystal of synthetic nuclear skyrmions.
	Remarkably, at the same time  the state  is an example of novel behavior of vortex matter
	being a crystal of linked closed vortex loops (see Fig.~\ref{SwedBread}a).
	
	In the solutions that we find the energy density is non-uniform and it breaks translation symmetry down to a three-dimensional crystal-like lattice.
	A typical unit cell of size $\lambda_{x,y,z}$ for the solutions we found is shown in Figure~\ref{SwedBread}a.
	Isosurfaces corresponding to constant value of the density of the first condensate ($|\Psi_1|\!=\!\textrm{const}$) represent a set of horizontal surfaces with an array of holes.
	Since it resembles several layers of Swedish bread ``kn\"{a}ckebr\"{o}d'', for brevity we will call it knackebrod phase. 
	While $|\Psi_2|\!=\!\textrm{const}$ isosurfaces are vertical cylinders passing through the knackebrod holes. 
	The phase winding of $\theta_2$ is equal to $\pm2\pi$ around each such a cylinder and  the signs ``+'' and ``-'' alternate in a checkerboard pattern.
	The phase $\theta_1$ has $\pm2\pi$ winding along the cross-sections of the knackebrod layer.
	Thereby, second component forms a lattice of vortex and anti-vortex lines.
	If one considers  the structure of only the second component, its cross-section is, in some respects, similar to spontaneous vortex-antivortex lattice in two-dimensional imbalanced chiral superconductor~\cite{barkman2018anti}.

	Now we establish the relationship between this solution and nuclear skyrmion crystal. 
	The important fact about the states that we find is that they do not contain points or domains in which  both order parameters are simultaneously zero. 
	In this regard, the topological density~ (\ref{charge}) can be calculated by assuming $\varphi_i\!=\!f_i/\sqrt{\sum_{j=1}^4f_j^2}$.
	Thereby, the integral over the density~(\ref{charge}) in both models gives the same index $Q$, also known as  skyrmion number. 
	Hence, we use the terms \textit{skyrmion}, \textit{antiskyrmion}, \textit{half-skyrmion} and \textit{quarter-skyrmion} to denote cases when $Q$ is 1, -1, 1/2 and 1/4 correspondingly.
	Obviously, in contrast to the mathematically idealised nuclear Skyrme model, in the model under consideration 
	the total index $Q$ is not an absolute invariant.
	The remarkable fact about the states of multicomponent imbalanced mixture that we find is that $Q$ does not change, unless perturbations are sufficiently strong.
	Hence it provides the important quantitative characteristic revealing the relationship with the nuclear skyrmions.

	The total skyrmion number $Q$ for the unit cell presented in Figure~\ref{SwedBread}a is four.
	This total $Q$ is composed of skyrmion numbers of each individual nodes -- positions with maximums of topological density~(\ref{charge}).
	When counting the nodes marked in orange in Figure~\ref{SwedBread}a, it is necessary to take into account that the same node may have several images due to periodic boundary conditions.
	Thus, we obtain that each node has $Q\!=\!1/2$ and hence is represented by a half-skyrmion.
	
	Note, that the unit cell has an internal symmetry and in fact consists of eight smaller rectangular cells which are identical in their energy density $\mathcal{F}(\mathbf{r})$ and topological index density $\rho_Q(\mathbf{r})$ distributions, but have different order parameter configuration. 
	We will call these $\mathcal{F}$-cells.
	
	Let us consider the distribution of order parameters in the vicinity of one half-skyrmion highlighted by rightmost black arrow. 
	The vertical dashed line passing through the center of the half-skyrmion depicts the vortex core in the second component.
	It  means that $\Psi_2\!=\!0$ along this line.
	Thick circle with an arrow denotes the path around which $\oint d{\theta_2}\!=\!2\pi$ while $|\Psi_2|$ and $\theta_1$ remain constant.
	Similarly to that, two parallel thick circles show two   loops around which $\oint d{\theta_1}\!=\!\pm 2\pi$. 
	Thus, closely spaced vortex loops are linked.
	
	One of the vortex cores in the first component ($\Psi_1\!=\!0$) is depicted by dashed loop. 
	Note, that cores of the second and first components (dashed curves) are also linked, since vertical line for the second component may be viewed as a part of a closed loop of infinite size. 
	Hence found phase can be considered as a collection of linked loops.
	But there is a subtlety: vertical vortex lines are passing through every hole in knackebrod -- white and black. 
	It reflects the fact that in the vicinity of points of linking we do not find compact skyrmions with $Q\!=\!1$, but rather pairs of half-skyrmions. 
	In other words, two adjacent holes of different color constitute one fractionalized $Q\!=\!1$ skyrmion. 
	This is an observation common for all solutions in the two-component fermionic model: the two complex fields form lattices consisting of  skyrmions fractionalized in half-skyrmions and even quarter-skyrmions.

	In the limit when the coupling between the components is negligible, or in single-component case, naturally there are no linkings.
	To highlight the difference with the two-component case, such a solution is presented in Figure~\ref{SwedBread}b, for $\gamma\!=\!0$.
	The structure of order parameter in the form of  parallel knackebrod layers is stable, and qualitatively has similar structure as in the two-component case.
	
\begin{figure*}
		\centering
		\includegraphics[width=0.7\linewidth]{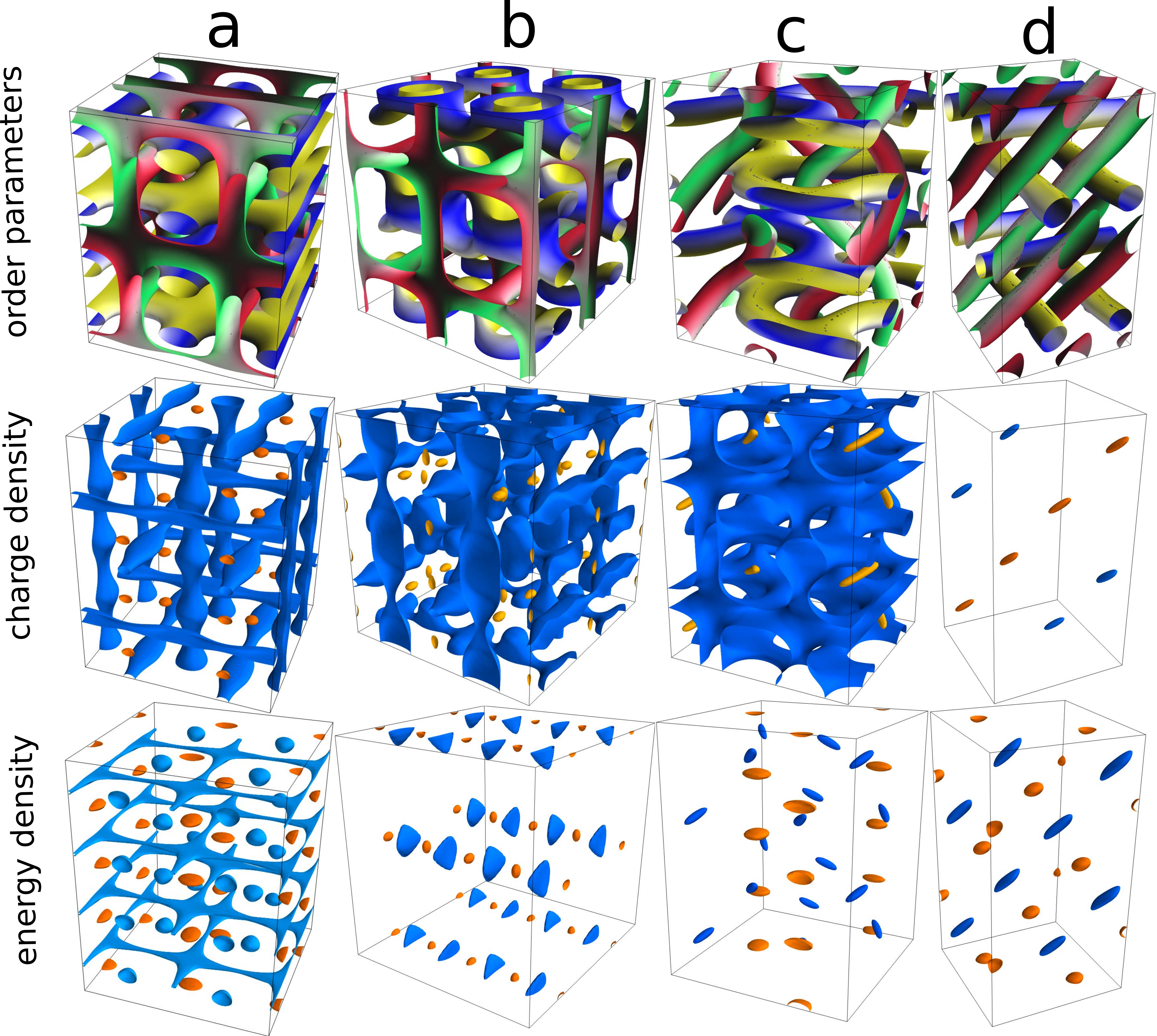}
		\caption{
			Synthetic skyrmion crystals of different symmetries.
			Top row shows isosurfaces of order parameters modulus for $|\Psi_{1,2}|\!=\!0.5\,\textrm{max}(|\Psi_{1,2}(\textbf{r})|)$, while color encodes the values of corresponding phases $\theta_{1,2}$ as in Figure~\ref{SwedBread}a.
			Middle row shows isosurfaces of the topological density: orange corresponds to $\rho_Q\!=\!0.95\,\textrm{max}(\rho_Q(\textbf{r}))$, for (\textbf{a})-(\textbf{c}) blue corresponds to $\rho_Q\!=\!0.05\,\textrm{max}(\rho_Q(\textbf{r}))$, for (\textbf{d}) -- to $\rho_Q\!=\!-0.95\,\textrm{max}(\rho_Q(\textbf{r}))$. 
			Bottom row shows isosurfaces of the energy density: blue corresponds to $\mathcal{F}\!=\!0.95\,\textrm{min}(\mathcal{F}(\textbf{r}))$, orange -- to $\mathcal{F}(\textbf{r})\!=\!0.95\,\textrm{max}(\mathcal{F}(\textbf{r}))$.
			\textbf{a} Unit cell of body-centered tetragonal (BCT) lattice with equilibrium periods $\lambda_{x,y}\!=\!15.1$, $\lambda_z\!=\!19.3$. Average energy density $\langle\mathcal{F}(\textbf{r})\rangle\!=\!-0.45$.
			\textbf{b} The different kind of stable BCT lattice characterized by the fact that half-skyrmions are fractionalized into quarter-skyrmions, $\lambda_{x,z}\!=\!21.0$, $\lambda_y\!=\!19.5$, $\langle\mathcal{F}(\textbf{r})\rangle\!=\!-0.44$.
			\textbf{c} DNA-like lattice characterized by energy density $\langle\mathcal{F}(\textbf{r})\rangle\!=\!-0.46$ and $\lambda_{x,y}\!=\!15.7$, $\lambda_z\!=\!20.7$. 
			For this solution the cores of half-skyrmions are noticeably elongated.
			\textbf{d} skyrmion-antiskyrmion lattice with $\langle\mathcal{F}(\textbf{r})\rangle\!=\!-0.46$, $\lambda_{x,y}\!=\!11.9$, $\lambda_z\!=\!21.9$.
		}
		\label{Km1_alpham1}
	\end{figure*}

	The Figure~\ref{SwedBread}a, 
	presents a simple tetragonal (ST) crystal. That is not the only possible state.
	We found a variety of solutions composed of half-skyrmions forming body-centered tetragonal (BCT) crystals.
	The family of stable states is depicted on Figure~\ref{Km1_alpham1}.  
	Let us classify these solutions according to their lattice type and morphology.
	
	Generalization of the ST solution appears when the system makes connections between vertical vortex lines of the second component.
	The stable solutions have these connections in the space between the knackebrod layers.
	Such solutions include twice as many layers in elementary cell. 
	The BCT crystal shown on Fig.~\ref{Km1_alpham1}a remains stable over  much larger range of parameters  than ST.
	By stability we mean that given crystal is protected by an energy barrier that prevents spontaneous transition to another crystal, expansion or shrinking of unit cell and other transformations.
	For this state index $Q\!=\!8$ per elementary cell. 
	Hence, each $\mathcal{F}$-cell has $Q\!=\!1$.
	It reflects the fact that this cell contains a pair of half-skyrmions.
	
	A very different stable state that the system forms can be viewed as twisted ST skyrmion lattice.
	This phase corresponds to   DNA-like configuration of vortices
	spanning the system shown on Fig.~\ref{Km1_alpham1}c.
	It consists of elongated half-skyrmions and has overall index $Q\!=\!8$  per elementary cell.  
	
	The system can form a crystal with more complicated fractionalization: composed of quarter-skyrmions.
	In such a crystal there are reconnections that occur between layers of the Knakebrod phase in BCT lattice. This gives chains of adjacent vortex loops, see Fig.~\ref{Km1_alpham1}b. We term this crystal BCT (chains) since it has skyrmion number $Q\!=\!8$.

	The synthetic nuclear Skyrme crystals presented above can be seen as spontaneous formation of lattices of skyrmion fractions with same $Q$. 
	However, we found solutions for which the signs of skyrmion number alternate. 
	An example of a stable state with total skyrmion number $Q\!=\!0$ is shown on Fig.~\ref{Km1_alpham1}d.
	This state, which isosurface morphology superficially resembles the structure of blue phases in liquid crystals~\cite{Wright_Mermin}, has an equal number of skyrmions and anti-skyrmions, i.e., it can be interpreted as a particle-antiparticle crystal.
	
	\section{Conclusions}
	
	The formation of crystal of topological excitations: vortex lattice, is considered
	a hallmark of superfluidity.
	A different kind of a lattice of topological solitons
	was explored in  the nuclear Skyrme model \cite{Skyrme}.
	There, substantial efforts were devoted to searching solutions   for nuclear skyrmion crystals
	that were hypothesized to describe matter in neutron star, which opened an
	active research direction in mathematical physics ~\cite{klebanov1985nuclear, goldhaber1987maximal, kugler1988new, castillejo1989dense}.
	An imbalanced Fermi superfluid is a seemingly unrelated system which is known to form well-studied FFLO state.
	By performing a numerical energy minimization of the Ginzburg-Landau model we demonstrated  that a multicomponent mixture of imbalanced superfluids has many stable states which are principally different from the FFLO state
	as well as from other known states such as  Larkin-Ovchinnikov crystals \cite{Bowers_Rajagopal}, interior gap \cite{liu2003interior} and Sarma phase \cite{sarma1963influence}, but is closely connected to the obsensibly unrelated Nuclear skyrmion crystals solutions \cite{klebanov1985nuclear, goldhaber1987maximal, kugler1988new, castillejo1989dense}. The states exist in a range of parameters and do not require a fine tuning.
	In the  states that we find nonlinear effects are highly important
	and structure of the states cannot be captured by naive ansatzes, but requires numerical solution of the full non-linear
	problem. 
	This calls for investigation whether these states are relevant in the microscopically different physics arising in dense QCD context \cite{Bowers_Rajagopal,casalbuoni2004inhomogeneous}.
	These states carry nontrivial density of topological index relating them to nuclear Skyrme crystals.
	Hence the system has spontaneous superflow in the form of a stable crystal of closed vortex loops. 
	This means that in the system with negative gradient terms, the vortex-antivortex lattice
	is a very competitive solution because it inherently has gradients both in the phases and densities of the fields

	Possible systems where nuclear Skyrme crystals may possibly be realized could be mixtures of $^6$Li and  $^{161}$Dy,
	$ {}^{171}$Yb and $ {}^{173}$Yb, $ {}^{161}$Dy and $ {}^{163}$Dy $ {}^{6}$Li and $ {}^{40}$K, $ {}^{167}$Er-$ {}^{161}$Dy 
	or mixture involving $ {}^{87}$Sr,  $ {}^{53}$Cr and  $ {}^{3}$He. 
	Fermi-fermi mixtures were experimentally explored for $ {}^{6}$Li and $ {}^{40}$K \cite{jag2016lifetime,naik2011feshbach} 
	Recently, the realization of $ {}^{40}$K-  ${}^{161}$Dy mixture was reported \cite{ravensbergen2019strongly}.
	The considered mixtures  have multiple stable states representing very different local minima.
	In order to make a crystal a global energy minimum, one may utilize the fact that their energy density is also a three-dimensional crystal. 
	I.e. when one adds an external, periodic in all three directions, potential $U(\textbf{r})|\Psi|^2$ these state can become a global minimum.  Likewise, axion-type additional terms~\cite{witten_axion}, which are proportional to topological density~(\ref{charge}) can make some of these states global minimum, since skyrmion crystals with skyrmion numbers of the same sign can get a significant energy gain.
	Even when the Skyrmion crystalls represent local minima,
	upon cooling the system would likely  
	form coexistent states or imperfect crystals, even in a box potential~\cite{box1,box2,box3}.
	The direct route to create these crystals in experiments is
	via a relaxation from an imprinted similar configuration. 
	There was recent progress in imprinting nontrivial topological charges
		\cite{lee2018synthetic}.
		Because the Skyrmion crystals can be viewed as a spontaneous lattice of
		vortex loops or lines, their experimental observation can be done via the same protocol
		as the observation of the ordinary vortex states.
		The other experimental route is the spectroscopy approach \cite{shin2007tomographic,schirotzek2008determination}.
		
		The nuclear Skyrme crystals that we find in context of cold atoms can form in unconventional superconductors where multicomponent models are ubiquitous.

		\begin{acknowledgments}
			We thank David Weld,  Martin Zwierlein, Alexander Zyuzin, Wolfgang Ketterle for useful discussions.
			We thank David Weld and  Martin Zwierlein for pointing out the possible experimental realizations of imbalanced fermionic mixtures with multicomponent order parameter.
			The work was supported by the Swedish Research
			Council Grants No. 642-2013-7837, 2016-06122, 2018-03659 and G\"{o}ran Gustafsson  Foundation  for  Research  in  Natural  Sciences  and  Medicine	and Olle Engkvists Stiftelse.
			A part of this work was performed at the Aspen Center for Physics, which is supported by National Science Foundation grant PHY-1607611.
		\end{acknowledgments}

	\section*{Appendix}
	\subsection{Parameters}
	
	In the simplest single component FFLO Ginzburg-Landau model \cite{Radzihovsky}, it is possible to rescale parameters so that all dimensional parameters are absorbed into $\alpha$, which is a function of  imbalance of fermionic populations and temperature. 
	To do so one should rescale: $\mathcal{F} \to \frac{\epsilon_F}{n \Delta_{BCS}} \mathcal{F}$, $x \to \frac{1}{q_0} x$ and $\Psi \to \Delta_{BCS} \Psi$. Where, according to \cite{Radzihovsky}, $q_0 = \frac{1.81 \Delta_{BCS}}{\hbar v_F}$, $\Delta_{BCS} = \frac{h_{c2}}{0.754}$ and $h_{c2}$ is a critical value for the chemical potential difference $h = \frac12 (\mu_\uparrow - \mu_\downarrow)$. Next, $\epsilon_F$ and $v_F$ are Fermi energy and velocity, $n$ is density of particles. Rescaled energy density becomes:
	\begin{equation}\label{ham_resc}
	\begin{split}
	\mathcal{F} = \zeta |\Delta\Psi|^2 + K |\nabla \Psi|^2  + \alpha |\Psi|^2 + \beta |\Psi|^4  + \\ +
	\xi \left(\mathrm{Im}[\Psi^* \nabla \Psi]\right)^2
	\end{split}
	\end{equation}
	
	with $\zeta = 0.61$, $K = -1.21$, $\alpha = 0.75 \ln(\frac{9 h}{4 h_{c2}})$, $\beta = 0.375$ and $\xi = 0.915$. Note, that $\alpha$ can be positive or negative, depending on the ratio $h / h_{c2}$.
	
	Two-component model necessarily has more parameters. We include the 
	bi-quadratic coupling term $\gamma|\Psi_1|^2 |\Psi_2|^2$. Apart from 
	that, when rescaling space $x \to \lambda x$ and energy density $\mathcal{F} \to h \mathcal{F}$, we have to multiply by the same scaling factor for both components. 
	It means that the model will depend on ratios of these factors for different components. Namely, space scaling factor for first component will be proportional to $\left(\frac{h_{c2}}{v_F}\right)_1 \left(\frac{v_F}{h_{c2}}\right)_2$, where indexes correspond to first and second components. Similarly for energy scaling we obtain factor $\left(\frac{n h_{c2}^2}{\epsilon_F}\right)_1 \left(\frac{\epsilon_F}{n h_{c2}^2}\right)_2$.
	
	In the mixtures of two species of fermionic ultracold atoms characteristics like densities will be different. It means that  
	in general we have $\zeta_1 \neq \zeta_2$, $K_1 \neq K_2$ etc. 
	The disparity in these parameters leads to a tendency to improve stability of the various  crystal solutions. Hence we set all parameters to be equal in our computations to assess the system in a regime that is {\it less} favorable for crystal formation. We take all coefficients to be of order unity and set $K_1 = K_2 \simeq -1$, $\alpha_1 = \alpha_2 = -1$, $\gamma = 0.5$ and other parameters equal to unity.
	
	\subsection{Numerical algorithm}
	
	We numerically minimized averaged energy density $\langle\mathcal{F}\rangle$ of the cuboidal unit cell of full three-dimensional model (see eq. (2) from main text) with periodic boundary conditions and rescaling of all 3 spatial sizes $\lambda_i$ of the cell independently.
	As a minimization routine we implemented nonlinear conjugate gradient algorithm, parallized on CUDA enabled GPU.
	The domain was discretisized by a mesh with $128^3$ points. 
	Some solutions were verified on the mesh with $256^3$ points. 
	In addition, we checked that the solutions remain stable on a relatively coarse grid with $64^3$ points.
	The second-order finite-difference discretization scheme was applied to continuous  Hamiltonian (see eq. (2) from main text).

	First of all, found solutions provide minimum of the energy for single unit cell of fixed size, i.e.   $E_0\!=\!\int\limits_{0}^{\lambda_x}\int\limits_{0}^{\lambda_y}\int\limits_{0}^{\lambda_z}\mathcal{F}(\mathrm{r})dxdydz\!\rightarrow\!\mathrm{min}$. Next, the periods $\lambda_{x,y,z}$ are in equilibrium, i.e. they are variables in the problem and were found such that simultaneously provide minimum for the average energy density, $\langle\mathcal{F}(\mathrm{r})\rangle\!=\!E_0/(\lambda_x\lambda_y\lambda_z)\!\rightarrow\!\mathrm{min}$.
	
	Similar approach, based on minimization of average energy density, was used to study one-dimensional helicoidal ordering in magnets without an inversion center~\cite{Dzyaloshinskii}. 
	Subsequent experiments confirmed the phenomena and the equilibrium periods on a quantitative level~\cite{Togawa_2012}.

	Because the system has many states representing local minima, to obtain a regular crystal
	one needs a suitable initial guess. As the initial guess for the
	numerical energy minimization we used first few terms of the Fourier series subjected to corresponding symmetry.
	Similar expressions were used in the search for nuclear Skyrme crystals \cite{kugler1988new}:
	$$
	\begin{cases}
	f_2 = \sum_{a,b,c}\alpha_{abc} \sin\left(\frac{2 \pi a x}{\lambda_x}\right) \cos\left(\frac{2 \pi b y}{\lambda_y}\right) \cos\left(\frac{2 \pi c z}{\lambda_z}\right)\\
	f_1 = \sum_{a,b,c}\beta_{abc} \cos\left(\frac{2 \pi a x}{\lambda_x}\right) \cos\left(\frac{2 \pi b y}{\lambda_y}\right) \cos\left(\frac{2 \pi c z}{\lambda_z}\right),
	\end{cases}
	$$
	
	and $f_3,\ f_4$ were found from $f_2$ by symmetry transformations.
	
	For example, for ST we used (for brevity we assume $\lambda_i = 2 \pi$) :
	$$
	\begin{cases}
	f_2 = \sin x,\ \ \ f_3 = \sin y,\ \ \ f_4 = \sin z\\
	f_1 = \cos x \cos y \cos z
	\end{cases}
	$$
	
	For BCT (chains):
	$$
	\begin{cases}
	f_2 = \sin x \cos y,\ \ \ f_3 = \sin y \cos z,\ \ \ f_4 = \sin z \cos x\\
	f_1 = c (\cos 2x + \cos 2y + \cos 2z) + (1 - 3 c) \cos 2x \cos 2y \cos 2z
	\end{cases}
	$$
	
	with $c \simeq 0.3$.

	\bibliographystyle{apsrev4-1}
%	\bibliography{fflo}

%

	\end{document}